\documentclass[12pt]{iopart}
\usepackage{iopams}
\expandafter\let\csname equation*\endcsname\relax
\expandafter\let\csname endequation*\endcsname\relax
\usepackage{amsmath}
\usepackage{amssymb}
\begin{document}
\title{Generalized Extensivity}
\author{John E. Gray$^1$, Stephen R. Addison$^2$\footnote{Corresponding Author}}
\address{$^1$Code B31,
Sensor Technology Branch,
Electromagnetic and Sensor Systems Department, 
Naval Surface Warfare Center Dahlgren, 18444 Frontage Road, Suite 328,
Dahlgren, VA 22448-5161}
\address{$^2$Department of Physics and Astronomy,
University of Central Arkansas,
201 Donaghey Avenue,
Conway, AR 72035-0001}
\ead{saddison@uca.edu}
\begin{abstract}
In order to apply thermodynamics to systems in which entropy is not extensive, it has become customary to define generalized entropies.  While this approach has been effective, it is not the only possible approach.  We suggest that some systems, including nanosystems, can be investigated by instead generalizing the concept of extensivity.  We begin by reexamining
the role of linearity in the definition of complex physical systems. We show that there is a
generalized form of extensivity that can be defined for a number of
non-linear systems. We further show that a generalization of the principle of linear
superposition is the basis for defining generalized extensivity.  We introduce a definition for the the degree of non-extensivity for systems. We show that generalized extensivity can be used as a means of understanding complex physical systems and we propose extending the idea extensivity beyond thermodynamics to other physical systems, including nanosystems.  
\end{abstract}
\pacs{05.20.-y, 02.50.Cw}
%\keywords{Tsallis Entropy, extensivity, generalized extensivity, nanothermodynamics}
%\submitto{\jpa}
\maketitle
\section{Introduction}
A number of authors have proposed generalizations of entropy to
address the problem of providing a universal measure of complexity or
disorder, examples include Tsallis entropy, and Renyi entropy \cite{Wacho2005}.   One of the properties of entropy that remains controversial is the issue of extensivity--the
appearance of linear or additive behavior of the subcomponents of a system
when taken as a whole. Extensivity is not a universal
feature of entropy, though it is often treated as if it is  \cite
{Callen1960,Callen1985}. However, a variety of sources have
demonstrated that entropic extensivity cannot be one of the fundamental
axioms of thermodynamics, see \cite{Addison2001} for further discussions.

To illustrate that extensivity is not a fundamental axiom of thermodynamics, we may consider a familiar example: entropy was defined phenomenologically by Clausius as an
integral over a reversible path,  but as Jaynes \cite{Jaynes1992} observed
the Clausius definition says nothing about extensivity, since the size
of the system does not vary over the integration path. Furthermore, Hill, in pioneering
work in the 1960s noted that extensivity does not apply to all entropic
systems, particularly those which exist at small scales \cite{Hill1962, Hill1987,Hill1994}. Thus, extensivity is a notion independent of entropy. The subject
of this paper is how to generalize extensivity so it applies to more general
complex systems. There is an advantage in attempting to generalize what can
be thought of as a purely mathematical concept. Extensivity can be generalized in a straightforward manner without bringing along the physical ideas associated with entropy; in this way we avoid the confusion that often arises from mixing the physical and mathematical aspects of entropy.

In complex, compound systems, the meaning of extensivity needs to  be explored in order to
gain insight into the meaning of complexity. Instead of proposing yet another
generalization of entropy, we will demonstrate the notion of extensivity can be generalized.  Generalized extensivity allows us to replace the classical
notion of extensivity by considering a more general notion of superposition.
By choosing to broaden the notion of what constitutes an extensive system
rather than changing the notion of what classical entropy means, we choose
to economize the physical interpretation of entropy;  this serves an economy of
purpose consistent with Occam's razor. 

We take a contrarian viewpoint to the proposed generalization of entropy by Tsallis \cite{Tsallis1988} as well as many subsequent ideas that are summarized in \cite{Tsallis2001}.
This approach allows us to suggest an approach to addressing the following the point made by Boon and Tsallis \cite{Boon2005}:
\begin{quote}
Asking whether the entropy of a system is or is not extensive \textit{without indicating the composition law of its elements}, is like asking whether some body is or is not in movement \textit{without indicating the referential with regard to which we are observing the velocity.}
\end{quote}
The formalism of \textit{generalized superposition} can be used to ask what
type of functions allow a nonlinear combination to give the appearance of
superposition.

We suggest several different forms of generalized extensivity can be defined, including one
associated with power laws which is connected to Tsallis entropy.  As part of
this exploration, we cast Tsallis entropy in a manner that is consistent
within the framework of generalized extensivity. Thus, instead of being
concerned whether Tsallis entropy  has the appearance
of extensivity or when it is inherently non-extensive, we propose that it
can be viewed as having the property of extensivity in the same context 
that a law of generalized superposition can be found, e.g. when it can be
related to power laws. We also discuss the role generalized superposition
plays in the understanding of power laws. 
\section{The Principle of Superposition and Generalized Extensivity}
The notion of extensivity can be developed mathematically by relating it to
the theory of linear systems. For a linear system with inputs $x_{i}(n)$,
scalars $c_{i}$ and a system transform $T\left( {}\right) $ there is always
a rule for combining inputs to give outputs. This system will be linear,
provided the system obeys the principle of superposition (where $+$ denotes ordinary
addition) 
\begin{equation}
T\left[ x_{1}(n)+x_{2}(n)\right] =T\left[ x_{1}(n)\right] +T\left[ x_{2}(n)
\right] 
\end{equation}
and scalar multiplication $\cdot $ 
\begin{equation}
T\left[ c\cdot \,x(n)\right] =c\,T\left[ x(n)\right] 
\end{equation}
There are many physical quantities where combinations behave
like linear systems and obey the principle of superposition. Mass and energy are examples of quantities that exhibit this property, when individual components of mass and energy are combined, the result is the combination of the individual masses and energies. Thus both mass and energy are extensive variables. On the other hand, both temperature and pressure are not extensive because they are not additive over sub-systems.

The most familiar example of a combination of components that is not
extensive is multiplication. For a system with a law of combination defined
by multiplication: 
\begin{equation}
x(n)=\left[ x_{1}(n)\right] \cdot \left[ x_{2}(n)\right] 
\end{equation}
with the introduction of a separator function for multiplication, the logarithm, 
\begin{align}
\log x(n)&=\log \left( \left[ x_{1}(n)\right] \cdot \left[ x_{2}(n)\right]
\right)\\& =\log \left[ x_{1}(n)\right] +\log \left[ x_{2}(n)\right].\notag
\end{align}
we have the appearance of superposition. For example, the Boltzmann
definition of entropy is $S=k\ln \Omega ,$ so under the circumstances that
the microstates are multiplicative $\Omega =\Omega _{1}\Omega _{2},$ then the
entropy of the combined states is $S=S_{1}+S_{2}$; so the Boltzmann entropy
obeys the principle of generalized extensivity. We then introduce the
definition 
\begin{equation}
S=\log [x(n)],
\end{equation}
so that Eq. (4) becomes 
\begin{equation}
S=S_{1}+S_{2}.
\end{equation}
This is an example of a non-linear combination of components that can
be re-interpreted as obeying superposition principle. This suggests that we
can look for an underlying generalized superposition principle, that is
extensivity, in many complex systems. Such a principle exists in signal
processing, in the analysis of homomorphic systems, that has wider applications in
complex systems as we now discuss.

\subsection{Generalized Extensivity}

In order to generalize the idea of superposition in linear systems to
non-linear systems, Oppenheim \cite{Oppenheim1967} proposed a modification
of Eq. (1) and Eq. (2). The summation sign in Eq. (1) must be replaced with two
different symbols: a rule for combining inputs designated by $\circledast $, and a rule for combining outputs designated by $\oplus $. We then
write the generalized superposition principle for such a system with a
system transform $H\left( {}\right) $ as 
\begin{equation}
H\left[ x_{1}(n)\circledast x_{2}(n)\right] =H\left[ x_{1}(n)\right] \oplus H
\left[ x_{2}(n)\right] 
\end{equation}
which is the generalization of Eq. (1) for nonlinear systems. A similar replacement is required for
scalar multiplication by a constant $c$, so the generalization of Eq. (2) is 
\begin{equation}
H\left[ c\,\otimes x(n)\right] =c\circ H\left[ x(n)\right] 
\end{equation}%
where $\otimes $ replaces the input scalar multiplication and $\circ $
replaces the output scalar multiplication. The notation we have adopted is and adaptation of the notation used in the
signal processing literature \cite{Oppenheim1975}.

Systems with inputs and outputs that satisfy both (7) and (8) are called 
\textit{homomorphic systems} since they can be represented as algebraically
linear (homomorphic) mappings between the input and output signal spaces.
Oppenheim and Schafer have written a brief history of signal processing
background for solving certain non-linear signal processing problems which
are the origin of their classification scheme  \cite{Oppenheim2004}.  They summarized this informally, by stating that all homomorphic systems can be thought of in the following manner:

\begin{quotation}
$\cdots$ all homomorphic systems  have canonical representation consisting of a
cascade of three systems. The first system is an invertible nonlinear
characteristic operator (system) that maps a nonadditive combination
operation such as a convolution into ordinary addition. The second system is
a linear system obeying additive superposition, and the third system is
the inverse of the first nonlinear system.
\end{quotation}

The logarithmic function is a separator function for multiplication. That
is why there is confusion about entropy being extensive, for many systems
the total entropy is achieved by the multiplication of subsystems, but this
is not always true. Hill \cite{Hill1962} includes examples where multiplication over subsystems does not  yield the true entropy of the composite system.

Most non-linear functions $N(x,y)$ do not separate with a superposition
principle, for example a separator function $S$ does not exist such that
\begin{equation}
S[N(x,y)]=A(x)+B(y).
\end{equation}
The mathematical details for determining if the rule for combining systems
can be deconstructed so that a valid separator function $S$ exists are discussed by Tretiak and Eisenstein 
\cite{Tretiak1976}. 

An informal argument is sufficient for our purposes to capture the flavor
of the conditions for $S$ to be a separator function. By taking the
partial derivative ($f_{,x}\triangleq \frac{\partial f}{\partial x}$) of Eq. (9)
with respect to each of the individual arguments, we have 
\begin{equation}
S^{\prime }[N(x,y)]N_{,x}(x,y)=A_{,x}(x),
\end{equation}
and 
\begin{equation}
S^{\prime }[N(x,y)]N_{,y}(x,y)=B_{,y}(y);
\end{equation}
combining we find 
\begin{equation}
\frac{N_{,x}(x,y)}{N_{,y}(x,y)}=\frac{A_{,x}(x)}{B_{,y}(y)}.
\end{equation}
Then, by defining 
\begin{equation}
H(x,y)=\ln \left( \frac{N_{,x}(x,y)}{N_{,y}(x,y)}\right) =\ln A_{,x}(x)-\ln
B_{,y}(y),
\end{equation}
we see that 
\begin{equation}
H_{,x}(x,y)=\frac{A_{,xx}(x)}{A_{,x}(x)},
\end{equation}
and 
\begin{equation}
H_{,y}(x,y)=-\frac{B_{,yy}(y)}{B_{,y}(y)};
\end{equation}
thus the first partials of $H(x,y)$ are functions of a single variable and
separation has been achieved. This provides the means of finding a
separator function for $N(x,y)$. In thermodynamic language, the existence of a separator function
means that there is a Maxwell relation between the
variables $A$ and $B$.

Generalized superposition provides an interpretation of the extensivity
property of entropy when it exists. In fact, generalized superposition can
be used as a guiding principle to look for extensive variables in a
generalized setting \cite{Gray1994, Gray1996} of determining
the thermodynamics of complex systems. More general types of entropy-like
variables can be defined using this form of generalized superposition. Since
the rule for combining subsystems into the whole is nonlinear, the principle
of generalized superposition allows us to look for a rule that gives the
appearance of linearity, and hence extensivity. Linearity on a macroscopic
scale gives us a generalized thermodynamics of complex systems. The
realization that the counting functions that enumerate the states have an
underlying\ separator function that obey this generalized form of
superposition allows us to generalize the concept of extensivity. Thus, we
define a system as obeying a principle of \textit{generalized extensivity} if
it obeys  Eqs. (7) and (9) rather than Eqs. (1) and (2).

\subsection{Other Forms of Extensivity}

A second possibility that allows us to consider another type of extensivity
is illustrated by considering a relaxation of the linearity requirement. An
enumeration function need not be strictly linear either. An example of this
is illustrated by attempting to linearize a counting function such as the
factorial $n!$. While $\ln (n!)$ is not linear, it is
approximately linear for a physical system that has a large number of
components; thus for all practical purposes
\begin{equation}
\ln ((n+m)!)\approx n\ln (n)+m\ln (m)\approx \ln (n!)+\ln (m!);\label{factorial}
\end{equation}
so $n!$ has been linearized with respect to the logarithm. We have omitted a factor $O(\frac{1}{n})$in Eq.\ref{factorial}, this term is
ignorable in systems that consist of a significant number of interacting
units. The only time this factor plays a role is when trying to determine
the transition point between collective and individual behavior, such as  in the
question of how many atoms are required before a system acts like a liquid
or solid, or at the nanoscale where collective behavior starts to break
down. (We will discuss this question further in a subsequent paper -- it is largely outside of the purview of this paper.) The difference
between approximately linear and linear is a point that is not emphasized as
examining the factorial function illustrates. There are a number of different functions
that are linear in the Stirling approximation. This notion of extensivity is captured by
what we term \textit{Ulam} \textit{extensivity} that is related to the
mathematical properties of functions that are termed approximately linear 
\cite{Hyers1945, Hyers1947, Hyers1952}.  Consider
\begin{equation}
f(x+y)-f(x)-f(y)\approx C+O(g(x)),
\end{equation}
where $g(x)$ tends to $0$ as $x$ becomes large,
which is the reason the logarithm of a factorial gives the appearance of
linearity for large $n$. Vogt \cite{Vogt1973} has discussed form isometry
which could be considered as a means to generalize Ulam extensivity.

\subsection{Power Laws, Homogeneity, and Extensivity}
Generalized extensivity connects power laws and homogeneity together. To
illustrate the concept of generalized extensivity we consider an example
of a system that has a power-law like combination rule for counting
combinations of states which is defined by 
\begin{equation}
x(n)=\left[ x_{1}(n)\right] ^{\alpha }\cdot \left[ x_{2}(n)\right] ^{\beta}
\end{equation}
where $\alpha$ and $\beta$ are constants constants. Notice that with the introduction of a
separator function for multiplication,the logarithm, we have 
\begin{align}
\log \left( x(n)\right)& =\log \left( \left[ x_{1}(n)\right] ^{\alpha }\cdot 
\left[ x_{2}(n)\right] ^{\beta}\right)\\&=\alpha \left( \log \left[ x_{1}(n)
\right]) +\beta(\log \left[ x_{2}(n)\right] \right).\notag
\end{align}
Thus recovering a familiar superposition principle.  If we then chose $\alpha=\beta$, we have
\begin{equation}
\log \left( x(n)\right)=\alpha \left( \log \left[ x_{1}(n)
\right] +\log \left[ x_{2}(n)\right] \right).\notag
\end{equation}
This equation reminds us of the basic definition of extensivity through the theory of homogeneous functions. We will now review the basic ideas through which we can analyze power laws using generalized extensivity.

The functional equations appropriate to the study of homogenous functions
were developed by Euler \cite{Euler1, Euler2}.  Davis \cite{Davis1962},
Stanley \cite{Stanley1971}, and Widder \cite{Widder} provide modern
introductions of varying degrees of sophistication to functional equations. In general, a function $f(x)$ is a homogenous function,
if for all values of the parameter $\lambda$,  
\begin{equation}
f(\lambda x)=g(\lambda )f(x)
\end{equation}
Stanley \cite{Stanley1971} has shown that $g$ is not arbitrary function, instead it is the parameter
raised to a power of $n.$ Thus a homogenous function $f(x$
) is one that satisfies 
\begin{equation}
f(x\lambda )=\lambda ^{n}f(x)
\end{equation}
This definition can be extended to any finite number of variables. The
degree, $n$, is restricted to integer values only. It is possible for
multidimensional functions to be homogenous of different degrees for different variables. This is common occurrence in thermodynamics. For a
function in the variables $x$, $y$, and $z$; the function satisfies 
\begin{equation}
f(\lambda x,\lambda y,z)=\lambda ^{n}f(x,y,z),
\end{equation}
we say this function is homogenous of degree $n$ in the variables $x$ and $y$
which it is not homogenous in the variable $z$. A definition
introduced by Stanley \cite{Stanley1971} of a generalized homogeneous function, is that it is a function that satisfies
the equation 
\begin{equation}
f(\lambda ^{a}x,\lambda ^{b}y)=\lambda f(x,y).
\end{equation}
It is this formulation of homogenous functions that is widely used in the
analysis of critical point phenomena and phase transitions using the static
scaling hypothesis. Functions that are homogenous obey the principle of
generalized superposition since $f\left( x\lambda \right) =\lambda ^{n}f(x)$
and $f\left( y\lambda \right) =\lambda ^{n}f(y)$ so in general $\lambda
^{n}f(x)+\lambda ^{n}f(y)\neq f\left( \lambda x+\lambda y\right) $, however
for $n=1$, we have 
\begin{align}
\lambda f(x)+\lambda f(y)&=f\left( x\lambda \right) +f\left( y\lambda \right)\\
&=f\left( \lambda x+\lambda y\right) =f\left( x+y\right), \notag
\end{align}
so homogeneity and extensivity are the same for some types of functions.
This form of extensivity is what we term \textit{homogenous extensivity}.

\subsection{Symbolic Extensivity}

The concept of \textit{symbolic extensivity} arises naturally from
dimensional analysis. Three approaches can be taken based on
the common aegis of dimensional analysis codified by Bridgman \cite
{Bridgman1922}. You take all the physical units that are required for a
complete system specification of a physical group of functions such as
length $L$, time $T$, mass $M$, etc.; \ so the dimensional symbols action
under the operations of arithmetic constitute a mathematical group. For
example, consider the symbol for length $L$. Combinations of $L$ form a
group $\mathcal{G}_{L}$: there is an identity, $L^{0}=1$; there is an
inverse, $1/L$ or $L^{-1}$; and $L$ raised to any rational power $p$ is a
member of $\mathcal{G}_{L}$ which has an inverse $L^{-p}$. The group
operation is multiplication, with the usual rules for handling exponents ($
L^{n}\times \ L^{m}=L^{n+m}$). For example in classical mechanics, any
physical quantity can be expressed dimensionally in terms of base units
which have dimensions $M$, $L,$ and $T$. 

From this group specification, by noting that by replacing a variable $x$ in 
$f\left( x\right) $ by $(x+\delta _{x})$, it is natural to define $f(x)$
to be \textit{translationally extensive} if 
\begin{equation}
f(x+\delta _{x})=f(x)+f(\delta _{x}).
\end{equation}
The \textit{degree of non-extensivity} $\Xi \left( x,\delta _{x}\right) $\
of a function $f(x)$ is then
\begin{equation}
\frac{f(x+\delta _{x})-f(x)-f(\delta _{x})}{f(x)}=\Xi \left( x,\delta
_{x}\right) ;
\end{equation}
so $\Xi \left( x,\delta _{x}\right) =0$ for an extensive function. For a
function that is Ulam extensive,
\begin{equation}
\Xi \left( x,\delta _{x}\right) =\frac{C}{g(x)}
\end{equation}
which is effectively zero for almost all $g(x)$, especially those that scale
with the number of objects. As an example, the area of square of side $x$
is $x^{2}$, so  $f(x)=x^{2}$, then $\Xi \left( x,\delta _{x}\right) =\frac{
\delta _{x}}{x}$, so for $\delta _{x}<<x$, $\Xi \left( x,\delta _{x}\right)
\simeq 0$ and area of square is approximately extensive; however, when $x$ becomes comparable in size to $\delta x$  the area is non-extensive.  This behavior is potentially observable at the transition region between macrosystems and nanosystems. To extend the
definition to multiple dimensions, we define the degree of non-extensivity
as  
\begin{align}
&\frac{f(x_{1}+\delta _{x_{1}},x_{2}+\delta _{x_{2}},...,x_{n}+\delta
_{x_{n}})-f(\overrightarrow{x})-f(\overrightarrow{\delta }_{\overrightarrow{x
}})}{f(\overrightarrow{x})}\\&=\Xi \left( \overrightarrow{x},\overrightarrow{
\delta }_{\overrightarrow{x}}\right).\notag
\end{align}
A multidimensional function is translationally extensive if $\Xi \left( 
\overrightarrow{x},\overrightarrow{\delta }_{\overrightarrow{x}}\right) =0$.
The volume of a cylinder is $\pi lr^{2}$, so the degree of non-extensivity is
\begin{align}
\Xi \left( r,\delta _{r},l,\delta _{l}\right) &=\frac{f(r+\delta
_{r},l+\delta _{l})-f(r,l)-f(\delta _{r},\delta _{l})}{f(r,l)}\notag \\&=2\frac{\delta
_{r}}{r}+\frac{\delta _{r}^{2}}{r^{2}}+\frac{\delta _{l}}{l}+2\frac{\delta
_{r}\delta _{l}}{lr}\notag
\end{align}
It is only when the fractional terms scale with the dimensional ratios, $
\frac{\delta _{r}}{r}$and $\frac{\delta _{l}}{l}$, that the degree of
non-extensivity is not effectively zero. For most volumes for which we have an
analytical formula, they can be translationally extensive until the dimensional
ratios, $\frac{\delta _{r}}{r}$and $\frac{\delta _{l}}{l}$, are of similar
size, then extensivity breaks down in a non-linear fashion.

The dimensional specification group, $\mathcal{G}_{S},$ is constituted from
the individual units of the dimensional quantities that constitute the units
of the overall complex system which are labeled $U_{1},U_{2},...,U_{n}$ in
the system specification which are specified by the groups $\mathcal{G}
_{U_{1}}$, $\mathcal{G}_{U_{2}}$,..., $\mathcal{G}_{U_{n}}$. The complex
system group is obtained by taking the direct product, $\otimes $, of these
groups :
\begin{equation}
\mathcal{G}_{S}=\mathcal{G}_{U_{1}}\otimes \mathcal{G}_{U_{2}}\otimes
...\otimes \mathcal{G}_{U_{n}}.
\end{equation}
It is possible to use the complex system group to formulate a dimensionless
specification for each complex system units provided they are specified by
scaling factors. Thus, for a complex physical system specified by variables $
x_{1}$, $x_{2}$, ...,$x_{n}$,\ can be represented as $P\left(
x_{1},x_{2},...,x_{n}\right) $. \ The way to remove the dimensions from a
physical variable is to replace $x_{1}\rightarrow \delta _{x_{1}}x_{1}$
where $\delta _{x_{1}}$has units proportional $\frac{1}{x_{1}}$, A function
that satisfies 
\begin{equation*}
f\left( \varepsilon _{x_{1}}x_{1}\right) =g\left( \varepsilon _{x_{1}}\right) f\left(
x_{1}\right) 
\end{equation*}
which is scaling or homogenous extensivity. Necessarily, since 
\begin{equation}
\lim_{\varepsilon _{x_{1}}\rightarrow 1}f\left( \varepsilon _{x_{1}}x_{1}\right) =f\left(
x_{1}\right) 
\end{equation}
which means $\lim_{\varepsilon _{x_{1}}\rightarrow 1}g\left( \epsilon _{x_{1}}\right) =1.$ So by an argument similar to the one used by Stanley, we have $g\sim x^{n}$. So this type of extensivity reduces essentially to
earlier work.
\section{Discussion and Conclusions}
The notion of extensivity extends beyond physics to the entire subject of
complex systems \cite{BarYam2003}. While the survey by Tsallis and Gell-Mann
discusses many non-physics applications of Tsallis entropy \cite{Gell-Mann2004}, we would argue that the same applies to generalized extensivity for much
the same reasons. Furthermore, extensivity and its generalization applies to
a new area of physics that is just starting to be explored:
nanothermodynamics, a name that has been proposed by Hill \cite{Hill2001a,Hill2001b}. Nanothermodynamics requires that chemical potential as
well as variables that are extensive require a reformulation by scaling them,
not relative to one scaling component $N$, but rather rescaling the multiple
components relative to scales of different sizes $N_{i}$. The appearance of
collective behavior, e.g. nanothermodynamics, emerges only if the
collective components act together in a unified manner, a type of
generalized extensivity \cite{Hill2001c}. 

In conclusion, we have proposed a means of extending the concept of extensivity to a wider
variety of physical systems as well as to the wider subject area of
complex systems. As a result of this, we suggest that the Tsallis
entropy could be interpreted as a form of generalized superposition for some
power laws.
\section{Acknowledgements}
The authors would like to thank Fransisco Santiago (NSWCDD) for being a
sounding board for discussion of new ideas to extend the concept of
generalized extensivity to the nanodomain.  The first author would like to thank Harold Szu for many useful discussions about the mathematical and physical aspects of entropy that helped form the author's attempt to separate the purely mathematical from the physical in understanding entropy.

\end{document}